\documentclass[11pt]{article}

\usepackage[english]{babel}
\usepackage[utf8]{inputenc}

\usepackage{algorithm,color}
\usepackage[intlimits]{amsmath} % necessary for \tag [] for limits in the inegral proper place
\usepackage{amsthm,mathrsfs}    % same
\usepackage[numbers]{natbib}
\usepackage{url}
\numberwithin{equation}{section}
\usepackage{amssymb,amsmath}
\usepackage{subfigure}
\usepackage{bm}
\usepackage{booktabs} % required for latex table
\usepackage{graphicx,graphics,epsfig}

\usepackage{float}
\usepackage{multirow}
\usepackage{array}

\theoremstyle{plain}
\newtheorem{thm}{Theorem}

\newcommand{\bthm}{\begin{thm}}
\newcommand{\ethm}{\end{thm}}

\newcommand{\bpf}{\begin{proof}}
\newcommand{\epf}{\end{proof}}

\theoremstyle{definition}

\usepackage[margin=.91in,a4paper]{geometry}
\usepackage[compact]{titlesec}
\titlespacing*{\section}{0pt}{1pt}{1pt}
\titlespacing*{\subsection}{0pt}{1pt}{1pt}
\titlespacing*{\subsubsection}{0pt}{1pt}{1pt}
\begin{document} 
\begin{center}
{\Large{\bf Statistics Educational Challenge in the 21st Century}}\\[.2in] %  A Technique Based on Superposition Principle
Subhadeep Mukhopadhyay\\
Department of Statistical Science,  Temple University\\ Philadelphia, Pennsylvania, 19122, U.S.A.\\[1.6em]
\end{center}
\begin{abstract}
What \emph{do} we teach and what \emph{should} we teach? An honest answer to this question is painful, very painful--what we teach lags decades behind what we practice. How can we reduce this `gap' to prepare a data science workforce of trained next-generation statisticians? This is a challenging open problem that requires many well-thought-out experiments before finding the secret sauce. My goal in this article is to lay out some basic principles and guidelines (rather than creating a pseudo-curriculum based on cherry-picked topics) to expedite this process for finding an `objective' solution.
\end{abstract}
\section{The Urgency of Statistics Education Reform}
At the 9th International Conference on Teaching Statistics (ICOTS9) meeting, Ronald L. Wasserstein, executive director of the ASA, in his talk ``Statistics in 2014: Reflections on the Occasion of the 175th anniversary of the American Statistical Association'' emphasized that education should be the top priority of Statistics profession. There is a growing need to reform the teaching of statistics to address the `Data Science Talent Gap' \citep{asa13b,asa13c,bigdnyt,bigdunc,D13USA}. I strongly believe statisticians in the 21st century (if they take advantage of their opportunities) can look forward to very bright futures for the discipline and profession of Statistics by solving the important problem of teaching thousands of aspiring statistical data scientist. Developing such a comprehensible introduction to comprehensive statistical modeling curriculum is challenging and requires radically \emph{new} approach.
\section{The Three Principles}
What should students learn to glimpse the frontiers of statistical theory and methods leading to Big Data science?  Interestingly, though, while there is a great deal of debate on the question `what \emph{should} we teach,' there is a general consensus on what we \emph{must} avoid. While discussing the challenges and opportunities for statistics education, in the next 25 years, \cite{kettenring2015} noted that ``There is a need to train students to use deep, broad, and creative statistical thinking instead of just training them in algorithms.'' A similar sentiment was echoed by the ASA 2014 Curriculum Guidelines for Undergraduate Programs in Statistical Science\footnote[2]{Available online: \url{http://www.amstat.org/education/pdfs/guidelines2014-11-15.pdf}}: ``\emph{Students need to see that the discipline of statistics is more than a collection of unrelated tools}.'' 

I summarize this as the following, which I call `\emph{The Exclusion Principle}':
\vskip.45em
{\bf Maxim 1}. We must \emph{not} design a Data Science training curriculum that looks like a long manual of specialized methods and series of cookbook algorithms. Otherwise, we will be in danger of producing DataRobots instead of Data Scientists \citep{deepblogDSDM}.
\vskip.45em

%students do not have to go through the struggle of bookkeeping and memorizing a long list of disconnected algorithms...We plan to exploit this new perspective in teaching to emphasize the underlying fundamental principles and statistical logic whose \emph{consequences} are algorithms.
The issue of what \emph{should} we teach was recently described aptly by David Donoho in his recent essay on ``50 Years of Data Science'' \citep{Donoho16}. This article proposed a comprehensive curriculum on Data Science (called `Greater Data Science'[\texttt{GDS}]) composed of six categories of activities: Data Exploration and Preparation, Data Representation and Transformation, Computing with Data, Data Modeling, Data Visualization and Presentation, and Science about Data Science. But the question remains:
\vskip.45em
{\bf Open Question}: \emph{How can we cover each of its 6 branches within a specified allotted time for the training program? How can we resolve this unsettled conundrum?}
\vskip.45em
As \cite{Donoho16} noted, `programs in Data Science cover only a fraction of the [\texttt{GDS}] territory'. The reason is very clear. To accommodate additional topics (like data pre-processing techniques, advanced computing, and other interdisciplinary real-data investigations) we run into the problem of {\bf `Too many topics, too little time.'} This is largely an open issue that presents unique challenges for 21-st century Statistics education. Thus, we first need to discover a new \emph{shortest} path (based on modern notations) to reach the \texttt{GDS} curriculum goal. This will require major shifts in \emph{how} (the language and notations) we teach.

In order to make way for more computer science–type materials, something must be sacrificed. But what should it be? This question was recently discussed by an expert panel (made up of a mix of data scientists, statisticians like David Hand, Chris Wiggins, Zoubin Ghahramani etc.) at Royal Statistics Society\footnote[2]{Available online: \url{https://www.youtube.com/watch?v=C1zMUjHOLr4}}. At the end, no consensus was reached on what tools and topics (small data techniques or modern large-scale high-dimensional techniques) to be included (or excluded) from the curriculum. Contrary to this current divergent and extreme approaches, I recommend an alternative philosophy to make both ends meet--`\emph{The Inclusion Principle}':
\vskip.45em
{\bf Maxim 2}. Teach methods for simple data in ways  \emph{that continue to work} for complex high dimensional data (similar to the goal of teaching finite-dimensional math in notations that extend to infinite-dimensional Hilbert space) -- A `\emph{comprehensive connected view}' of statistical data modeling.
\vskip.45em
This will bypass the problem of ``depth vs. breadth'' and at the same time could provide the students a glimpse to the frontiers of statistical theory and methods. To further accelerate, and enhance students' learning, I advocate an additional principle --\emph{`Principle of Parsimony'}, which says
\vskip.4em
{\bf Maxim 3}. Prefer the education program that covers the curriculum using a \emph{minimum number of fundamental tools, concepts, and notations}.
\vskip.45em
Can we develop a new curriculum based on these three postulates, thereby strengthening the statistical core for data science among students and applied researchers? Yes, I believe. The only question is when.
\section{Intrinsic vs. Extrinsic Principles}
Extrinsic operational principles aim to answer the question of the following type: Theory first or Application first? edX, Coursera or distance learning? R, Python or Matlab? Shiny, GoogleVis, or Plotly? R Markdown or Leaflet? All of these are important questions. However, this is not what I think will inspire students to become a statistical scientist, at the first place. We have to distinguish between IT-training and Science education.
Before sexifying the subject we need to pay attention how to unify using basic fundamental principles that will simplify the practice and accelerate students learning.

\vskip.5em
I welcome comments from readers about the views expressed in this article. Please take a moment to complete this short survey: \url{https://www.surveymonkey.com/r/FKRXHM7}.

\end{document}